\documentclass[namedreferences]{solarphysics}
\usepackage[hyperref,optionalrh,solaromanenum]{spr-sola-addons} 
\usepackage{graphicx}                    
\usepackage{color}                       
\usepackage{breakurl}                         

\newcommand{\degr}{\ensuremath{^\circ}~}

\newcommand{\arcsec}{\ensuremath{^{\prime\prime}}~}
\begin{document}

\begin{article}

\begin{opening}

\title{The Temperature -- Magnetic Field Relation in Observed and Simulated Sunspots}

%
\author[addressref={aff1},corref,email={michal.sobotka@asu.cas.cz}]{\inits{M.}\fnm{Michal}~\lnm{Sobotka}} 
\author[addressref={aff2},email={rrezaei@iac.es}]{\inits{R.}\fnm{Reza}~\lnm{Rezaei}} 

%
\runningauthor{M. Sobotka, R. Rezaei}
\runningtitle{Temperature -- Magnetic Field Relation in Sunspots}

\address[id={aff1}]{Astronomical Institute of the Czech Academy of Sciences (v.v.i.),
   Fri\v{c}ova 298, 25165 Ond\v{r}ejov, Czech Republic}
\address[id={aff2}]{Instituto de Astrof\'\i sica de Canarias, v\'\i a L\'{a}ctea,
   38200 La Laguna, Tenerife, Spain}

\begin{abstract}
Observations of a relation between continuum intensity and magnetic field strength in sunspots have been made during nearly five decades. This work presents full-Stokes measurements of the full-split ($g = 3$) line Fe {\sc i} 1564.85 nm with spatial resolution of 0.5\arcsec obtained with the {\it GREGOR Infrared Spectrograph} in three large sunspots. The continuum intensity is corrected for instrumental scattered light and the brightness temperature is calculated. Magnetic field strength and inclination are derived directly from the line split and the ratio of Stokes components. The continuum intensity (temperature) relations to the field strength are studied separately in the umbra, light bridges, and penumbra. The results are consistent with previous studies 
and it was found that the scatter of values in the relations increases with increasing spatial resolution thanks to resolved fine structures. The observed relations show trends common for the umbra, light bridges, and the inner penumbra, while the outer penumbra has a weaker magnetic field compared to the inner penumbra at equal continuum intensities. This fact can be interpreted in terms of the interlocking comb magnetic structure of the penumbra. A comparison with data obtained from numerical simulations was made. The simulated data have a generally stronger magnetic field and a weaker continuum intensity than the observations, which may be explained by stray light and limited spatial resolution of the observations and by photometric inaccuracies of the simulations.
\end{abstract}

%

\end{opening}

%
\section{Introduction}
\label{s:int} 

The inhibition of convection by magnetic field in sunspots is an obvious reason to expect that the temperature and magnetic field are coupled. The magnetic field magnitude $B$ and continuum intensity $I_{\rm c}$, from which the brightness temperature $T_{\rm b}$ can be derived, were measured at different positions in one or more sunspots by several authors. \cite{Abdus71} first made a scatter plot of $I_{\rm c}$ versus $B$, based in observations of the spectral line Fe {\sc i} 630.25 nm. \cite{GurHo81} claimed to obtain a linear relation between $I_{\rm c}$ and $B$ but did not attempt to interpret it. \citeauthor{Marti90} (\citeyear{Marti90}, \citeyear {Marti93}) suggested to study the relation $T_{\rm b} \sim B^2$ relevant to the horizontal pressure balance in sunspots. \cite{Kopp92} for the first time utilised the infrared line Fe {\sc i} 1564.85 nm with Land\'e ~factor $g = 3$ to study this relation and found that it was linear only in the umbra but not in the whole sunspot.

The line Fe {\sc i} 1564.85 nm seems to be very suitable to study the coupling of $I_{\rm c}$ or $T_{\rm b}$ with $B$. The line is only moderately sensitive to the temperature and it is formed in the low photosphere at the height of about $h = 110$ km above $\tau_{500} = 1$ \citep{Bruls91}, while a nearby continuum is formed at $h = -30$ km \citep{Verna81}. It splits completely already at $B = 500$ G and larger field-strength magnitudes are simply proportional to the wavelength difference between the peaks of its Zeeman $\sigma$-components \citep{Solan92}. In spite of the fact that the line suffers from some blends in the umbra, which, together with broadened wings at low temperatures, reduces the accuracy of the measured umbral field strength \citep{Rueed95}, it can be used to measure $B$ directly without a need of line-inversion methods. Another advantage is that in the infrared, the scattered light, caused by the telescope and instrument, and also the image degradation by seeing is lower than in the visible region.

In the works that followed, the infrared line Fe {\sc i} 1564.85 nm was used to derive magnetic information either from direct measurements of the $I$ and/or $V$ Stokes profiles \citep{Livin02,Penn03,Reza12} or by means of line-inversion methods \citep{Solan93,Mathe04,Jaegg12}. Other authors \citep{Balth93,Stanc97,Leona08} utilised spectral lines in the visible region and line-inversion methods. Scatter plots of $B$ versus $I_{\rm c}$ or $T_{\rm b}$ derived from measurements including the umbra together with the penumbra \citep{Kopp92,Solan93,Balth93,Stanc97,Mathe04} have a typical non-linear shape: an umbral part with increasing $B$ at lowest intensities (temperatures), an umbra-penumbra transition with a relatively small change of $B$ in a wide intensity range, and a penumbral part characterised by a substantial decrease of $B$.
A recent survey of numerous spots made by \cite{Jaegg12} presents a similar picture. The authors interpret the non-linear $B - T_{\rm b}$  relation in the umbra as a consequence of molecular (H$_2$) formation, reducing the number density of particles and the gas pressure in the coolest parts, so that a stronger magnetic field is necessary to maintain the horizontal pressure balance.

We extend the previous studies by high spatial resolution full-Stokes observations that newly make it possible to study the temperature -- magnetic field relation separately in the umbra, light bridges, and the penumbra. We also take advantage of existing radiative MHD numerical simulations of sunspots (\citealp{Rempe12}, in particular) and compare our results with synthetic data.

\section{Observations and Data Analysis}
\label{s:oda}

\subsection{Observations}
\label{ss:obs}

The data were acquired on 11 May 2015, from 09:30 to 11:30 UT with the {\it GREGOR Infrared Spectrograph} (GRIS, \citealp{Colla12}) attached to the 1.5 m solar telescope {\it GREGOR} \citep{Schmi12}. The telescope is equipped with an adaptive optics system \citep{Berke12}, which corrects aberrations due to the seeing and telescope, and an image rotator, compensating the solar image rotation caused by the {\it GREGOR's} alt-azimuth mount. GRIS full-Stokes scans in the spectral region centred at the Fe {\sc i} 1564.85 nm line were obtained by moving the spectrograph slit by 300--400 steps 0.13\arcsec wide, which is also the resolution element of GRIS along the slit. The wavelength sampling in this region was 3.95 pm, the exposure time 100 ms, and each polarization state was accumulated three times to increase the signal-to-noise ratio. The polarimetric calibration was made using the telescope model and the {\it GREGOR} polarimetric calibration unit \citep{Hofma12}. The continuum intensity was measured in a line-free region at $\lambda = 1564.23$ nm.

Our target was the large and complex active region NOAA 12339, which was composed of four large sunspots (two leading and two following) and many smaller spots. It was located near the disc centre and started to decay at the day of observations. We selected three best scans for further analysis: 007 (09:31--09:47 UT, leading spot 1), 016 (10:51--11:03 UT, leading spot 2), and 018 (11:10--11.24 UT, following spot 1). The heliocentric angle of all the spots was equal to 15\degr ($\mu = 0.97$) at the times when their scans were acquired. The field of view of the scanned maps varied from 61\arcsec $\times$~52\arcsec to 61\arcsec $\times$~39\arcsec according to the number of scanning steps.

During the observations, the adaptive optics worked throughout all scans, indicating a seeing characterised by the Fried parameter between $r_0 = 10$--15~cm. Because the wavefront sensor worked at $\lambda = 500$ nm and $r_0 \sim \lambda^{6/5}$, the Fried parameter at $\lambda = 1565$ nm was approximately 40--60~cm. The spatial resolution of 0.5\arcsec was estimated from the smallest features seen in the continuum map.

\subsection{Scattered Light}
\label{ss:sca}

The observed intensity suffers from parasitic light scattered in the atmosphere, optical system of the telescope, and spectrograph. The influence of the atmosphere in the near infrared is small and its major part is compensated by the adaptive optics. The spectral scattered light ({\it i.e.} in the $\lambda$-direction) introduced by the spectrograph mainly influences line depths in Stokes $I$ and it can be neglected for $Q,\, U,\, V$ because their continua are at zero. Generally, the effect of spectral scattered light is not important for measurements of line splitting, shifts, and $I$-continuum intensity, and we can neglect it in our analysis.

The most important is the spatial scattered light in the telescope and spectrograph. Its 2D point-spread function (PSF) was determined for {\it GREGOR} and GRIS under seeing conditions similar to ours by \cite{Borre16}. It was composed of two Gaussian functions, which described the scattering from narrow and wide angles:
   \begin{equation}  \label{e:psf}
     {\rm PSF}(x,y) = p_{\rm n}G_{\rm n}(x,y,\sigma_{\rm n}) + p_{\rm w}G_{\rm w}(x,y,\sigma_{\rm w}) \,.
   \end{equation}
Parameters of the wide-angle term $p_{\rm w} = 0.2$ and $\sigma_{\rm w} = 20\arcsec$ were derived under the assumption that when the magnetic-field vector {\textit {\textbf B}} is parallel with the line of sight (hereafter LOS), which occurs at some places inside the umbra, the observed central component of the full-split line Fe~{\sc i} 1564.85~nm is formed completely by scattering of light from the surrounding penumbra and granulation. This term corresponds to scattered light in the telescope's optics. Parameters of the narrow-angle term $p_{\rm w} = 0.8$ and $\sigma_{\rm w} = 0.18\arcsec$ were derived from a spectral scan of a pinhole array inserted in the science focus of the telescope. This term corresponds to the scatter inside the spectrograph.

We deconvolved the observed $I$-continuum maps with this PSF in the Fourier domain, using the Wiener optimum restoration filter $F$ ({\it e.g.} \citealp{Pratt78}) in the form
   \begin{equation}  \label{e:wie}
     F(\omega) = \frac{C+1}{C} \frac{{\rm MTF}(\omega)}{{\rm MTF}^2(\omega) + 1/C} \,,
   \end{equation}
where MTF (modulation transfer function) is the Fourier image of PSF, $\omega$ is the spatial frequency in a radially symmetric MTF, and $C = 30$ is a parameter depending on the signal-to-noise ratio.
An increase of this parameter raises the contribution of high spatial frequencies in the restoration, including a noise. Its value was found empirically following the rules that i) no part of real information was suppressed and ii) the noise was not enhanced.
This way we obtained the deconvolved maps of $I_{\rm c}$ for all scans. 

\subsection{Brightness Temperature}
\label{ss:tem}

The brightness temperature $T_{\rm b}$ can be easily obtained from the deconvolved $I_{\rm c}$ maps normalised to the mean continuum intensity of undisturbed photospheric granulation, using the Planck function \citep{Solan93,Mathe04}. Following these authors, we adopt the reference quiet-Sun temperature equal to 7058 K. This value corresponds to the temperature at $\tau_{1600} = 1$ in the photospheric reference model of \cite{Maltb86}. Due to the opacity minimum at 1600 nm, this optical depth refers to deeper and hotter layers than $\tau_{500} = 1$. The obtained function $T_{\rm b}(I_{\rm c})$ for the continuum at $\lambda = 1564.2$ nm is approximately linear in the $I_{\rm c}$ range from 0.5 to 1.1.

\subsection{Magnetic Field Measurement}
\label{ss:mag}

The Fe {\sc i} 1564.85 nm normal Zeeman triplet shows a complete split for $B > 500$~G \citep{Solan92}. We can use the well-known formula
   \begin{equation}  \label{e:bee}
     B = \frac{\Delta \lambda}{4.67\times 10^{-5} \, g\lambda_0^2} \  \   {\rm [G, cm]}
   \end{equation}
to calculate the magnetic field strength magnitude from the displacement $\Delta \lambda$ of $\sigma$-components from the central wavelength $\lambda_0$. To reduce the effect of noise in the spectra as much as possible, we measure the line split $s = 2 \Delta \lambda$ using a 7-point parabolic fit to the extrema in the Stokes profiles $\overline{QU} = \sqrt{Q^2 + U^2}$ and/or $V$ as follows:

We calculate the total circular and linear polarizations $P_{\rm cir} = \int |V(\lambda)| \, {\rm d}\lambda$ 
and $P_{\rm lin} = \int  \overline{QU}(\lambda) \, {\rm d}\lambda$ 
in a 0.35~nm wide region around the central wavelength.
Then the splits $s_{V}$ and/or $s_{\,\overline{QU}}$ are measured from the $V$ and/or $\overline{QU}$ profiles only when $P_{\rm cir}$ and/or $P_{\rm lin}$ are higher than 0.5, respectively. The total polarization values serve as weights in the final calculation of the split $s$, so that
   \begin{equation}  \label{e:spl}
     \Delta \lambda = s/2 = \frac{P_{\rm cir} \, s_{V} + {P_{\rm lin} \, s_{\,\overline{QU}}}}{2(P_{\rm cir} + P_{\rm lin})} \,.
   \end{equation}
One resolution element in $\Delta \lambda$ corresponds to 58 G and it is further refined by the parabolic fit to the extrema.

The magnetic field inclination $\gamma$ related to the LOS direction can be determined directly as 
$\gamma = \arctan(\sqrt{Q_{\rm max}^2 + U_{\rm max}^2}/V_{\rm max})$. According to \cite{Solan92}, the ratio 
$\sqrt{Q_{\rm max}^2 + U_{\rm max}^2}/V_{\rm max}$ 
for the line Fe {\sc i} 1564.85 nm depends only on $\gamma$ and not $B$ for $B > 1000$ G outside the umbra and $B > 2500$ G in the umbra. We adopted a conservative condition of $B > 1500$ G for the $\gamma$ calculations, being aware of a possible underestimate of $\gamma$ in the umbra. The fact that the $Q, \, U, \, V$ profiles are not corrected for scattered light is another source of uncertainty, leading to a possible overestimate of $\gamma$. Thus we have to consider the LOS inclination values only as a rough estimate, particularly in the umbra.

\section{Results}
\label{s:res} 

%
\begin{figure} 
\centerline{\includegraphics[width=1.0\textwidth]{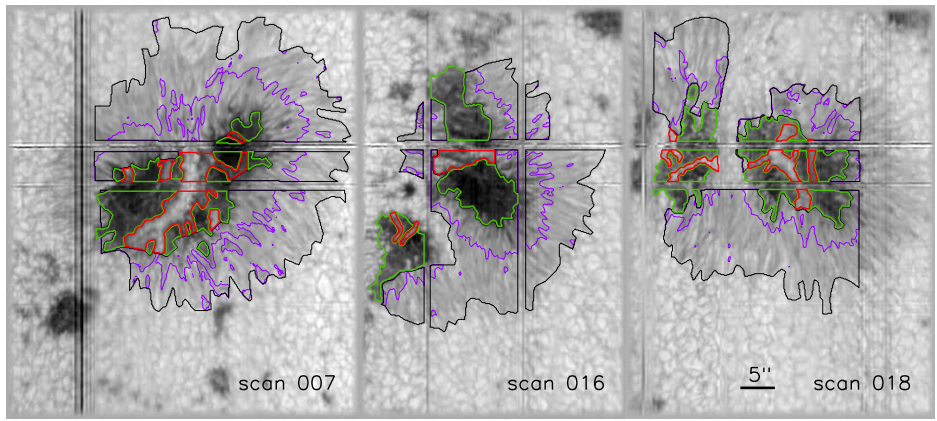}}
\caption{Continuum intensity maps, corrected for scattered light, obtained from the scans 007, 016, and 018. Contours delineate three types of masks that distinguish the umbra (green), light bridges (red), and the penumbra (black). The purple line is a contour of $B = 1400$ G, considered as an approximate boundary between the inner and outer penumbra. The vertical size of images is 61\arcsec.}
\label{f:masks}
\end{figure}

Regions of interest, which are the umbra, light bridges, and penumbra, were defined by binary masks that were drawn by hand using the $I_{\rm c}$ maps corrected for scattered light (Figure~\ref{f:masks}). These masks also served to exclude problematic areas caused by scanning errors and dirts on the spectrograph slit.

\begin{figure} 
\centerline{\includegraphics[width=1.0\textwidth]{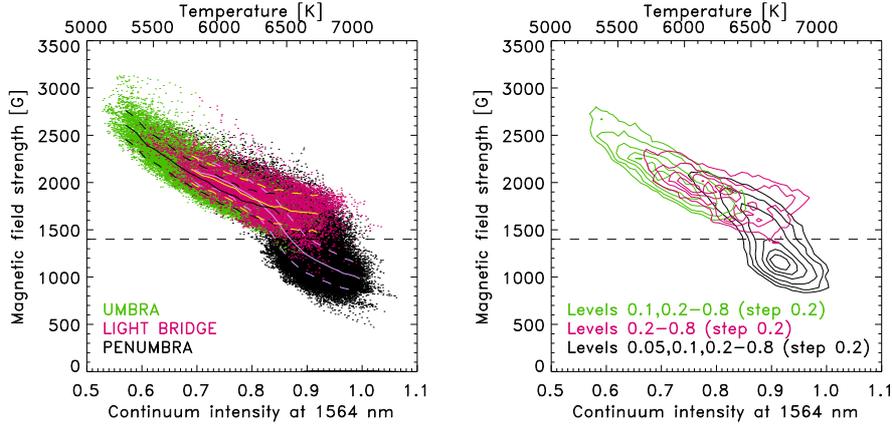}}
\caption{{\it Left:} Scatter plot of $B$ versus $I_{\rm c}$ and $T_{\rm b}$ in the umbra (green, 27826 points), light bridges (red, 9901 points), and the penumbra (black, 113218 points) for all three sunspots together. Solid lines show average values in the umbra (black), light bridges (yellow), and penumbra (purple), together with dashed lines of $\pm 1\sigma$, which characterise the scatter of individual points. {\it Right:}  Density contours of the scatter plot. Density values are normalised to their maxima and contour levels are annotated in the plot. The horizontal dashed line marks $B = 1400$ G (cf. Figure~\ref{f:masks}).}
\label{f:biall}
\end{figure}

Scatter plots of $B$ versus $I_{\rm c}$ and $T_{\rm b}$ for all three spots and all regions of interest together are shown in Figure~\ref{f:biall}, left panel. The points, corresponding to different positions in the fields of view, are green in umbrae, red in light bridges, and black in penumbrae. Scatter plots that were made separately for individual sunspots (not shown) have a very similar shape. Solid lines represent mean values of $B$ at points with $I_{\rm c}$ falling into 0.01 wide bins of the continuum intensity histograms. The bins must contain at least 200 points to calculate the mean value. We also computed standard deviations $\sigma$ characterising the scatter of $B$ of individual points in each bin (dashed lines). Colours of the lines are black for the umbra, yellow for light bridges, and purple for penumbra.
Because the ranges of the umbra, light bridges, and penumbra partially overlap, contours of relative densities of points in the scatter plots are depicted for a better clarity in the right panel of the figure. The relative densities are normalised to their maxima.
Compared to the previous studies, our results have a substantially larger scatter due to resolved fine structures; our spatial resolution is at least twice better than that of \cite{Mathe04}.

In the umbra, the magnetic field strength varies between 1600 and 3100 G with a typical scatter of $\pm 150$ G. The normalised continuum intensity (temperature) ranges from 0.53 (5100 K) to 0.85 (6400 K) and the typical scatter is $\pm 200$~K, caused by numerous umbral dots. The $B - I_{\rm c}$ relation is definitely non-linear and its curvature even increases in a corresponding $B^2 - T_{\rm b}$ plot (not shown).
A possible reason for the deviation from linearity was discussed by \cite{Jaegg12}. Our results are consistent with those of \cite{Kopp92}, who made measurements at various positions inside six sunspots. \cite{Livin02}, \cite{Reza12}, and \cite{Watson14} concentrated to the darkest points in multiple umbrae. Our results are in a good agreement with the first work and $B$ measured by us in the whole umbra is weaker by 200--300~G compared to the two latter ones.
A compilation of results published by different authors can be seen in Figure~7 of \cite{Penn03}.

Light bridges,  with magnetic field strength between 1300 and 2500~G and continuum intensity (temperature) from 0.62 (5500 K) to 0.97 (6900 K), show a $B - I_{\rm c}$ relation very similar to that of the umbra, slightly shifted to higher continuum intensities and temperatures. The scatter of points is larger than in the umbra, $\pm 200$ G in $B$ and $\pm~250$ K in $T$.

%

The $B - I_{\rm c}$ relation in the penumbra is more complex.
The density plot in Figure~\ref{f:biall} shows a dense cloud of points below $B = 1400$~G and an extension towards higher magnetic field strengths (1400--2300~G) and lower continuum intensities and temperatures (0.70--0.97, 5800--6900~K). The contour $B = 1400$~G plotted in Figure~\ref{f:masks} indicates that this value may be considered as an approximate boundary between the inner and outer penumbra. The cloud characterised by a weaker magnetic field (700--1400~G) and higher continuum intensity and temperature (0.83--1.01, 6400--7100~K) corresponds to the outer penumbra. The magnetic field strength rises quite steeply towards the inner penumbra, while the temperature at the continuum formation height diminishes only slightly. Below $I_{\rm c} < 0.83$ in the inner penumbra, the $B - I_{\rm c}$ relation is similar to that of the umbra and it is nearly identical with that of light bridges (Figure~\ref{f:biall}, left panel).
The large scatter of intensities and temperatures is obviously due to the presence of bright and dark penumbral filaments.

%
\begin{figure} 
\centerline{\includegraphics[width=1.0\textwidth]{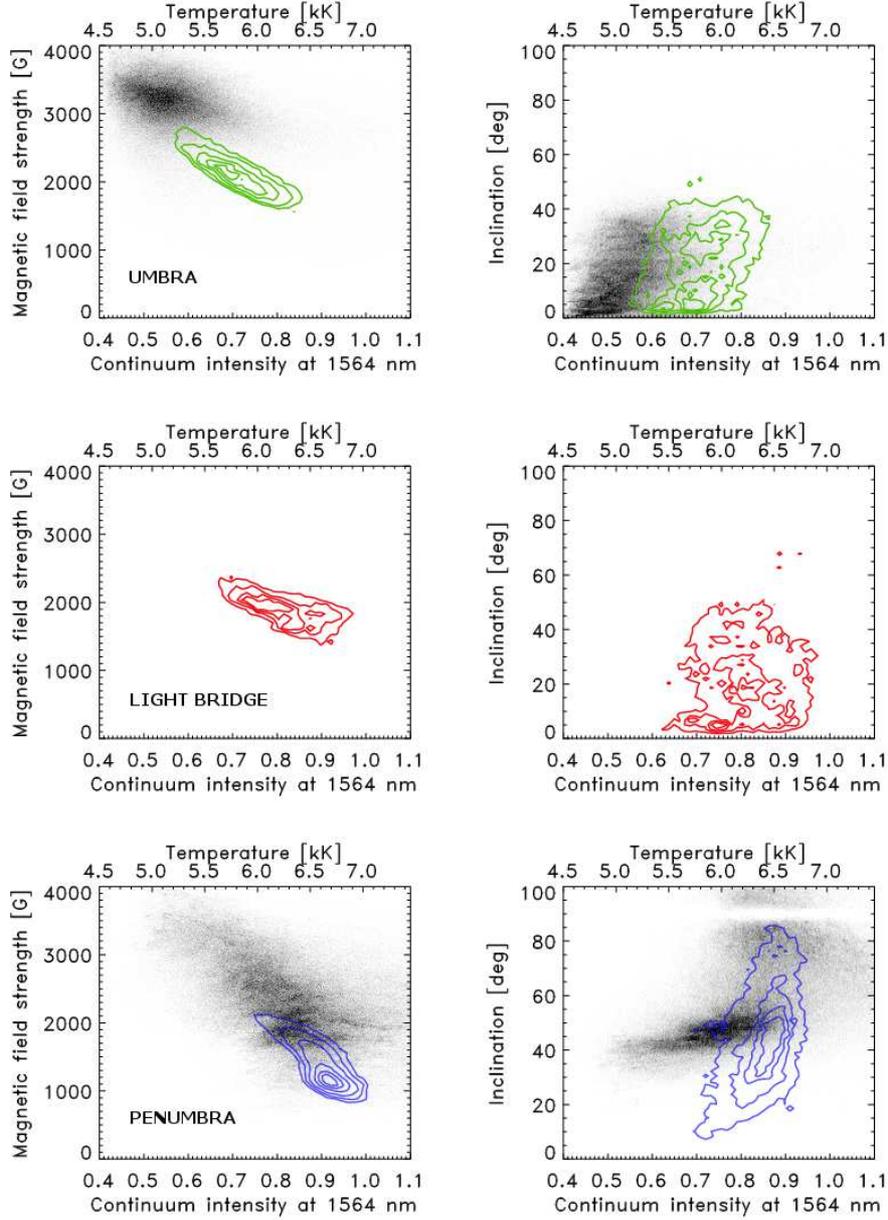}}
\caption{Scatter-plot densities of $B$ versus $I_{\rm c}$ and $T_{\rm b}$ ({\it left}) and $\gamma$ versus $I_{\rm c}$ and $T_{\rm b}$ ({\it right}), separately for the umbra ({\it top}), light bridges ({\it middle}), and the penumbra ({\it bottom}). 
The contours represent observed data for all three spots, while the grey-scale clouds depict in a linear scale the relative densities of scatter plots obtained from numerical simulations. The contour levels are as in Figure~\ref{f:biall} in the left column and 0.1--0.7, step 0.2 in the right column.
Note that $\gamma$ was measured only in regions where $B > 1500$ G.}
\label{f:psep}
\end{figure}

Figure~\ref{f:psep} shows separate density plots of the umbra, light bridges, and penumbra for all three sunspots together. The left column depicts density contours of the $B - I_{\rm c}(T_{\rm b})$ relation identical to those in Figure~\ref{f:biall} and the right column presents density contours of magnetic field LOS inclination $\gamma$ versus $I_{\rm c}$. The plots include 27826 measured points in the umbra, 9901 in light bridges, and 113218 in the penumbra. Because $\gamma$ was measured only in the locations with $B > 1500$~G, the number of points in the inclination plots was reduced to 9374 in light bridges and 31749 in the penumbra, so that the whole outer penumbra is missing.

Note that the $\gamma - I_{\rm c}$ scatter-plot densities were obtained for the LOS reference angle equal to 15\degr from the normal.
The umbra has a large scatter of points but the density plot shows an increase of $\gamma$ with $I_{\rm c}$, which is an effect of increasing field inclination towards the edge of the umbra and in umbral dots \citep{Socas04}.
The LOS inclination in light bridges is similar to that in the umbra. A major part of the points is scattered between 0--50\degr and the average is around $20{^\circ}$. \cite{Jurca06} found a higher value of $\gamma = 40$\degr connected with a magnetic canopy above light bridges at $h \approx 200$~km above $\tau_{500} =1$. The line Fe~{\sc i} 1564.85 nm, however, is formed 100~km below this canopy, where the magnetic field vector can still be almost vertical.
In the inner penumbra, the individual values are spread over 10--90\degr similarly to the results of \cite{Mathe04}.

\section{Comparison with a Synthetic Sunspot}
\label{s:syn}

It is interesting to compare our observational results with analogous data derived from numerical simulations of a sunspot, free of instrumental effects, scattered light, spectral line blends, and providing a much higher spatial resolution. We used the numerical model of a sunspot located at the disc centre \citep{Rempe12}, particularly {\tt slab\_12x8x12km\_ng}\footnote{See {\tt http://download.hao.ucar.edu/pub/rempel/sunspot\_models/Fine\_Structure/}.}, which simulates a 6144 km wide strip of the umbra, penumbra, and surrounding granulation with the horizontal sampling of $12 \times 12$~km.
This data set was derived by M. Rempel from a grey radiative transfer sunspot simulation at originally $16\times 16\times 12$ km resolution in a $49.152\times  49.152\times 6.144$ Mm ($x, y, z$) domain, evolving for one hour from an initial state defined by other models with lower resolution. The top half of that domain was regridded to $12\times 12\times 8$ km resolution and evolved for another 15 minutes (10 minutes grey and the last 5 minutes non-grey).
A snapshot of this model was utilised by \cite{Borre14} to compute full-Stokes synthetic spectra in the 1.5 $\mu$m region, employing the synthesis module of the SIR code (Stokes Inversion based on Response functions; \citealp{Basil92}).
These forward modeled spectra were used for comparison with our observations.

Splits of the synthetic $V$ and $\overline{QU}$ profiles of the Fe~{\sc i} 1564.85~nm line as well as the LOS inclination were measured in the same way as in the observations (Section~\ref{ss:mag}). The umbra and penumbra were distinguished by binary masks drawn by hand in the synthetic continuum map. Light bridges do not exist in the synthetic sunspot. Relations between the magnetic field strength/inclination and continuum intensity/temperature based on the synthetic data are displayed in Figure~\ref{f:psep}
in the form of grey-scale density plots. Scales of the plots are linear, from zero to the maximum density.

We can see from the figure that in the umbra, synthetic $B$ is generally higher by approximately 1000~G, $I_{\rm c}$ lower by 0.15, and the scatter of points is larger than in the observations. In the penumbra, the differences are smaller but still significant: 700~G, 0.08, and the scatter is substantially larger.
Atomic and molecular blends may reduce the observed $B$ in the umbra and the light scattered in the Earth's atmosphere (we corrected only for the instrumental stray light) can raise the observed continuum intensity. Moreover, highly magnetic features obtained by numerical simulations in the umbra and penumbra increase the simulated $B$ but they are too small to be detected in observations with the resolution of 350~km.

Also the sunspot simulation has some photometric limits. According to the notes included for the Rempel's models, ``the luminosity of the quiet Sun surrounding the sunspot is not exactly solar'', being off by a few percent.  It is possible that the value used for the quiet-Sun intensity in the model is spuriously increased by bright magnetic elements in this highly resolved simulation, which reduces the normalised intensity. There is also a question of how accurate are the intensities of this model, since the model evolved only for one hour at this resolution and the non-grey radiative transfer was switched on only during the last five minutes.

Realising that the spatial resolution in the simulations is much better than in the observations, we suggest that the scatter of points increases with increasing spatial resolution. This is seen already when we compare our results with previous works that have a lower resolution. The simulations are fifteen times finer than our observations and the corresponding scatter is much larger. It can be considered that the $B - I_{\rm c}$ scatter is an intrinsic attribute of sunspot fine structures.

It is hard to compare the observed LOS magnetic field inclination with the local-reference-frame inclination in the simulations. However, because the LOS angle is only 15\degr in our observations, a tentative comparison may be interesting. In the umbra, both clouds of points are quite similar when we omit the differences in $I_{\rm c}$ caused by the scattered light in our observations and possible photometric inaccuracies of the simulations. The trend of increasing inclination with increasing $I_{\rm c}$ is common for both observed and synthetic data. In the penumbra, the points with $\gamma < 30$\degr are absent in the simulations. These points appear in the observations probably due to the effect of projection into the LOS system. We also have to keep in mind that the inclination was measured only in the inner penumbra.

\section{Discussion and Conclusions}
\label{s:dis} 

We repeat the classical exercise of finding the temperature -- magnetic field relation in sunspots in the infrared region, using the simplest direct methods to measure the magnetic field strength and inclination. Our results are consistent with the previous works, showing the typical shape of the $B - T_{\rm b}$ relation (see Section~\ref{s:int}). Thanks to high spatial resolution provided by {\it GREGOR}, we are able to treat the umbra, penumbra, and light bridges separately.
While the authors of previous works, {\it e.g.} \cite{Stanc97}, \cite{Mathe04}, and \cite{Jaegg12}, utilised an intensity threshold to separate the umbra and penumbra, we used manually drawn boundaries based on visible structures. Thanks to this morphological criterion we obtained overlaps in the properties of the umbra, penumbra, and light bridges. All these structures have a common overlap in the range $1700\, {\rm G}<B< 2100\, {\rm G}$ and \mbox{$5800\, {\rm K}<T_{\rm b}<6300\, {\rm K}$} (see Figure~\ref{f:biall}). In our results, this range corresponds to the ``umbra-penumbra transition'' in the typical $B - T_{\rm b}$ plot.

Despite the large scatter caused by fluctuations of temperature and magnetic field strength in fine-scale structures, which is also typical for the synthetic sunspot, there are general trends in the $B - T_{\rm b}$ relation represented by the mean values plotted in Figure~\ref{f:biall}. These trends are very similar in the umbra, light bridges, and the inner penumbra in the range of the common overlap. This indicates that the interaction of the magnetic field with moving plasma, which is the cause of sunspot cooling, has the same character in the inner parts of spots, where the magnetic field is not strongly inclined to the normal. We find a tentative separation value $B = 1400$ G between the inner and outer penumbra. The outer penumbra shows a generally weaker magnetic field strength compared to the inner penumbra at equal temperatures.

These facts can be explained in terms of the ``interlocking comb'' magnetic structure of the penumbra, (\citealp{ThoWe08}, Chapter 5.2, and references therein), where at least two magnetic-field systems are expected. The first one, which can be considered as a continuation of the umbral magnetic field, has less inclined field lines that extend high into the atmosphere. 
Its magnetic field is stronger and its filamentary structure (spines) weakens in the outer penumbra. In the inner penumbra, it is related to dark filaments \citep{Borre11}. The second system, which is connected with the Evershed flow, has a weaker magnetic field with steeply inclined, nearly horizontal field lines and a limited vertical extent. The corresponding penumbral filaments are bright in the inner penumbra and dark in the outer one. The largest intensity contrast between the bright and dark penumbral filaments is seen in the inner penumbra close to the umbral border, where bright penumbral grains \citep{Mulle73} are formed at the tips of the bright filaments.

When observed in the low photosphere (at $h \approx 100$ km), the umbra, light bridges, and inner penumbra are dominated by the first,
strong, and less inclined magnetic field system. Intensity differences in the inner penumbra are large due to the high contrast of penumbral filaments. The outer penumbra, however, is dominated by the nearly horizontal magnetic field at this height and the field strength is considerably weaker because i) it decreases with increasing distance from the umbra and ii) the second field system is generally weaker than the first one \citep{Langh05}.
The contrast of penumbral filaments is lower than in the vicinity of the umbra, so that the intensity differences are smaller than in the inner penumbra.

The statistical analysis of the temperature -- magnetic field relation is complementary to the state-of-art methods of spectropolarimetric inversions and numerical simulations, which are currently used to study sunspots. It can serve to verify their results and even to bring some new details. Our approach made it possible to compare individual $B - T_{\rm b}$ relations obtained for different sunspot structures and we find a strong similarity between the umbra, light bridges, and inner penumbra, while the outer penumbra with $B < 1400$ G shows a different behaviour. This can be explained by the interlocking comb magnetic structure. There are still some questions open: Do other sunspots with different size, brightness, and phase of evolution show a different $B - T_{\rm b}$ relationship? And does it vary with the solar cycle?

%
\begin{acks}
This work was supported by the grant 14-04338S of the Czech Science Foundation, the FP-7 Capacities Project No. 312495 SOLARNET, and the institutional support RVO:67985815 of the Czech Academy of Sciences. R.R. acknowledges financial support by the Spanish Ministry of Economy and Competitiveness through the project AYA2014-60476-P.
We thank J.~M. Borrero for synthetic spectra computed in the frame of the international working group Extracting Information from Spectropolarimetric Observations: Comparison of Inversion Codes at the International Space Science Institute (ISSI) in Bern (Switzerland).
We use data provided by M. Rempel at the National Center for Atmospheric Research (NCAR). The National Center for Atmospheric Research is sponsored by the National Science Foundation.
The 1.5-meter {\it GREGOR} solar telescope was built by a German consortium under the leadership of the Kiepenheuer Institute for Solar Physics in Freiburg with the Leibniz Institute for Astrophysics Potsdam, the Institute of Astrophysics G\"ottingen, and the Max Planck Institute for Solar System Research in G\"ottingen as partners, and with contributions by the Instituto de Astrof\'isica de Canarias and the Astronomical Institute of the Czech Academy of Sciences. We thank the referee for comments leading to a substantial improvement of the paper.

\noindent {\bf Conflict of Interest:} The authors declare that they have no conflict of interest.
\end{acks}


%
\bibliographystyle{spr-mp-sola}
\bibliography{TBspots}  

\end{article} 
\end{document}